\documentclass[prd,onecolumn,notitlepage,nofootinbib,longbibliography,superscriptaddress,floatfix]{revtex4-2}

\usepackage{amsmath,amssymb}
\usepackage[colorlinks=true, allcolors=blue]{hyperref}
\usepackage{graphicx,placeins}
\usepackage[italicdiff]{physics}
\usepackage{comment}
\usepackage{booktabs}
\usepackage{xcolor}

\newcommand{\Dn}{\Delta_n}
\newcommand{\Eth}{E_n^\mathrm{th}}
\newcommand{\be}{\begin{equation}}
\newcommand{\ee}{\end{equation}}

\makeatletter
\newcommand{\twocolclearpage}{%
  \close@column@grid
  \clearpage
  \twocolumngrid
}
\makeatother

\begin{document}

\title{Superluminal constraints from ultra-high-energy neutrino events}

\author{J.M. Carmona}
\email{jcarmona@unizar.es}
\affiliation{Departamento de Física Teórica and Centro de Astropartículas y Física de Altas Energías (CAPA), Universidad de Zaragoza, Zaragoza 50009, Spain}

\author{J.L. Cortés}
\email{cortes@unizar.es}
\affiliation{Departamento de Física Teórica and Centro de Astropartículas y Física de Altas Energías (CAPA), Universidad de Zaragoza, Zaragoza 50009, Spain}

\author{M.A. Reyes}
\email{mkreyes@unizar.es}
\affiliation{Departamento de Física Teórica and Centro de Astropartículas y Física de Altas Energías (CAPA), Universidad de Zaragoza, Zaragoza 50009, Spain}

\begin{abstract}
The $220^{+570}_{-100}\,$PeV neutrino detected by KM3NeT marks the beginning of ultra-high-energy neutrino astronomy and provides a powerful probe of Lorentz Invariance Violation (LIV). In superluminal scenarios, neutrinos can decay through vacuum $e^-e^+$ pair emission or neutrino splitting. Previous analyses of the KM3-230213A event relied on simplified survival-probability estimates and, in some cases, used inaccurate decay-width expressions or neglected redshift and threshold effects. In this work we present a unified and self-consistent framework that corrects these issues and applies to both the energy-independent ($n=0$) and quadratic ($n=2$) superluminal cases. We collect and recast the decay-width and threshold expressions, clarify their flavor dependence, and include a consistent treatment of cosmological propagation. We also assess the impact of cascade regeneration and show that cascade effects are negligible for the purpose of setting LIV bounds. The survival-probability approximation adopted in previous works is therefore justified, while our framework provides a coherent basis for future analyses of superluminal neutrino constraints, which should consistently include possible time-delay signatures.
\end{abstract}

\maketitle

\section{Introduction}

The recent detection of a neutrino event with an energy of $220^{+570}_{-100}$\,PeV by KM3NeT~\cite{KM3NeT:2025npi} has opened the window of ultra-high-energy (UHE) neutrino astronomy. While its origin—whether it arises from a point source, the diffuse background, or is a cosmogonic neutrino—remains uncertain~\cite{Neronov:2025jfj,KM3NeT:2025aps,Li:2025tqf,Dzhatdoev:2025sdi,Das:2025vqd,Boxi:2025ony}, its mere observation confirms the existence of a population of neutrinos surpassing the 100\,PeV threshold~\cite{Muzio:2025gbr}. This detection also raises intriguing questions, particularly about the absence of similar events in IceCube, which has so far only reported neutrinos up to an order of magnitude lower in energy~\cite{KM3NeT:2025ccp}. However, it is reasonable to anticipate that this event is just the first of many within this energy regime, to be observed by IceCube, KM3NeT, and next-generation UHE neutrino observatories such as IceCube-Gen2 (radio), RNO-G, GRAND, and POEMMA, which will significantly increase the exposure to neutrinos in the EeV range.

This newly opened window provides an unprecedented opportunity to explore new physics. In fact, the particularities of the KM3-230213A event have been used to connect it with phenomena as diverse as oscillations and sterile neutrinos~\cite{Brdar:2025azm}, dark matter~\cite{Borah:2025igh,Aloisio:2025nts,Dev:2025czz,Kohri:2025bsn}, and primordial black holes~\cite{Boccia:2025hpm,Anchordoqui:2025xug,Airoldi:2025opo,Airoldi:2025bgr,Zantedeschi:2024ram}. 

A population of UHE neutrinos have general implications for new interactions and neutrino decays. In particular, it affects scenarios of Lorentz Invariance Violation (LIV), where neutrinos with a modified dispersion relation may undergo decay during propagation, altering the expected flux of cosmic neutrinos~\cite{Stecker:2014oxa}. 

Indeed, a few studies have already placed LIV constraints on superluminal neutrinos using the KM3-230213A event~\cite{KM3NeT:2025mfl,Yang:2025kfr,Satunin:2025uui}. They have, however, some limitations. The analysis in~\cite{KM3NeT:2025mfl} considers a restricted scenario where the neutrino velocity is assumed to be independent of the energy, neglects the expansion of the universe, and does not take properly into account the threshold energy of the decay process. The study in~\cite{Yang:2025kfr} focuses exclusively on the so-called quadratic scenario, under the assumption that the detected neutrino is part of the diffuse flux, with a Monte Carlo approach relying on formulas that are not entirely accurate. On the other hand, the constraints derived in ref.~\cite{Satunin:2025uui} for the quadratic scenario are specific to the detection of this particular neutrino, and assume some arbitrary choices for the distance to its source. Moreover, the analyses of refs.~\cite{KM3NeT:2025mfl,Satunin:2025uui} are based on the probability of survival of the detected event, without considering that the observed neutrino could, in principle, be the product of previous neutrino decays. One may therefore wonder about the validity of this approximation and about the possible influence of cascade effects on the inferred limits.

The present work revisits these analyses within a unified and self-consistent framework. We incorporate a proper treatment of the decay thresholds, the energy and flavor dependence of the decay widths, and the cosmological propagation of neutrinos. We also examine the validity and limitations of the survival-probability approach by explicitly studying the potential influence of cascade regeneration on the detected flux. This provides a consistent methodology applicable to both energy-independent and energy-dependent superluminal scenarios, and clarifies the domain of applicability of current and future constraints based on ultra-high-energy neutrino observations.

The structure of the paper is as follows. In section~\ref{sec:thdecay} we review the expressions of the decay widths relevant for different LIV scenarios, emphasizing their unified formulation and comparing them with those adopted in the literature. These results are then applied in section~\ref{sec:simplified} to derive constraints on superluminal neutrinos using the survival-probability method, incorporating the proper redshift dependence and decay thresholds. In section~\ref{sec:cascades} we examine the possible impact of cascade regeneration on these results and demonstrate that its contribution is negligible for setting LIV bounds, thereby justifying the validity of the simplified approach. Section~\ref{sec:KM3future} applies this framework to the KM3-230213A event and discusses its implications for future ultra-high-energy neutrino observations, including the connection with possible time-delay effects. Finally, section~\ref{sec:conclusions} summarizes our conclusions and outlines prospects for future studies.

\section{Unification of decay widths for superluminal neutrinos}
\label{sec:thdecay}

Superluminal neutrinos attracted worldwide attention after the initial OPERA claim on superluminal propagation in 2011~\cite{OPERA:2011ijqv1}. Soon after, Cohen and Glashow noted~\cite{Cohen:2011hx} that a superluminal neutrino of high enough energy should quickly lose energy by decaying to an electron-positron pair, $\nu_\alpha \to \nu_\alpha e^- e^+$, where $\alpha$ represents the flavor of the decaying neutrino (electron, muon or tau). They gave a first estimate of the neutrino decay width in a model with an energy-independent velocity, $v=1+\delta/2$, coming from a relation between energy and the modulus of the momentum
\begin{equation}
    E^2 \,=\, \vec{p}^2 \left(1+ \delta\right)\,,
    \label{MDRn0}
\end{equation}
where the neutrino mass has been neglected. More generally, one may consider an energy-dependent superluminal behavior, negligible at low energies but increasingly relevant at high energies. The dispersion relation would be in that case
\begin{equation}
    E^2 \,=\, \vec{p}^2 \left[1+ \left(\frac{|\vec p|}{\Lambda}\right)^n\right]\,,
    \label{MDRn12}
\end{equation}
where we also neglect the neutrino mass. A dispersion relation like eq.~\eqref{MDRn12} gives a momentum-dependent superluminal velocity $v=1+((n+1)/2)\left(|\vec p|/\Lambda\right)^n$, where usually only the linear ($n=1$) and quadratic ($n=2$) cases are considered. Within an LIV Lagrangian framework, in the $n=1$ case, the superluminal energy-momentum relation, eq.~\eqref{MDRn12}, would only apply either to the neutrino or to the antineutrino (while the other would be subluminal). Instead, in the $n=2$ case, if the neutrino is superluminal, this will also be the case for the antineutrino~\cite{Carmona:2022dtp}. 

We will refer as $n=0$ to the case of an energy-independent superluminal velocity, corresponding to eq.~\eqref{MDRn0}. We can then unify the notation for the $n=0$ and $n\neq 0$ cases by writing
\begin{gather}
    E^2=\vec{p}^2\left[ 1+ |\vec p|^n \Dn\right],\\
    v^2=1+(n+1)E^n\Dn,\\
    \Delta_0\doteq \delta, \quad \Delta_{n> 0} \doteq \Lambda^{-n}\,.
\end{gather}

More accurate results for the decay width were obtained in refs.~\cite{Carmona:2012tp,Bezrukov:2011qn}, which showed that the original Cohen-Glashow estimate~\cite{Cohen:2011hx} follows from a prescription that cannot be derived from a consistent Lagrangian framework. More recently, ref.~\cite{Carmona:2022dtp} improved the calculation of both total and partial widths by incorporating the --- previously neglected --- charged-current contribution to the vacuum $e^-e^+$ pair-emission (VPE) channel, and by considering that another decay process, $\nu_\alpha\to \nu_\alpha\nu_\beta\bar{\nu}_\beta$, neutrino splitting (NSpl), is allowed for $n\neq 0$. The decay widths for the processes $(i)=\{\text{VPE},\text{NSpl}\}$ can be written as
\begin{equation}
    \Gamma_{\alpha,n}^{(i)}(E)\simeq K_{\alpha,n}^{(i)}\,\frac{G_F^2}{192\pi^3}\,E^{5+3n}\,\Dn^3\,,
    \label{eq:decaywidth}
\end{equation}
where the $\simeq$ symbol means that $E$ is well above the threshold of the corresponding process. The values of the constants $K_{\alpha,n}^{(i)}$ are given in tables~\ref{table:Ks_VPE} and \ref{table:Ks_NSpl}, and take into account that neutrino splitting can proceed through three channels, corresponding to each possible flavor $\beta$ of the neutrino-antineutrino pair.\footnote{The results in tables~\ref{table:Ks_VPE} and \ref{table:Ks_NSpl} are taken from ref.~\cite{Carmona:2022dtp}. Note, however, that in order to follow the usual convention in the $n=0$ case, where $\delta$ is defined as $\delta = v^2 - 1$, the parameter $\Lambda$ in the present work is defined through the relation between $E^2$ and $\vec{p}^{\,2}$, whereas ref.~\cite{Carmona:2022dtp} used the relation between $E$ and $|\vec{p}|$. Consequently, the high-energy scale in ref.~\cite{Carmona:2022dtp} corresponds to $2^{1/n}\,\Lambda$.\label{footnote1}}

\begin{table}[tbp]
\centering
\begin{tabular}{cccc} \toprule
     \hspace{4em} &  $\alpha=\mu,\tau$ & $\alpha=e$ & Averaged \\ \midrule
     $n=0$ & 17/420 & 221/420 & 17/84\,(0.202) \\
     $n=1$ & 121/1680 & 1573/1680 & 121/336\,(0.360) \\
     $n=2$ & 81/910 & 81/70 & 81/182\,(0.445) \\ \bottomrule
\end{tabular}
\caption{Values of the constants $K_{\alpha,n}^{(i)}$ for the VPE in the $n=0, 1$ and 2 cases. The subindex $\alpha$ stands for the flavor of the decaying neutrino: $e$, $\mu$ or $\tau$. These values have been obtained by approximating the square of the sine of the Weinberg angle by $1/4$.}
\label{table:Ks_VPE}
\end{table}

\begin{table}[tbp]
\centering
\begin{tabular}{ccc} \toprule
     \hspace{4em} &  $\alpha=\mu,\tau,e$ & Averaged \\ \midrule
     $n=0$ & 0 & 0 \\
     $n=1$ & 22/75 & 22/75\,(0.293) \\
     $n=2$ & 1422/5005 & 1422/5005\,(0.284) \\ \bottomrule
\end{tabular}
\caption{Values of the constants $K_{\alpha,n}^{(i)}$ for the NSpl in the $n=0, 1$ and 2 cases. The subindex $\alpha$ stands for the flavor of the decaying neutrino: $e$, $\mu$ or $\tau$. These values have been obtained by approximating the square of the sine of the Weinberg angle by $1/4$.}
\label{table:Ks_NSpl}
\end{table}

The decay through VPE has a kinematical threshold
\begin{equation}
    \Eth =  \left(4 m_e^2 \,\Dn^{-1}\right)^{1/(n+2)}\,
    \label{eq:Eth}
\end{equation}
where $m_e$ is the electron mass. Due to the small neutrino mass, the NSpl threshold is much lower than for VPE and will not affect the following discussion. We shall therefore treat separately the regimes above and below the VPE threshold. For $E>\Eth$, the total decay width must account for both decay channels,
\begin{equation}
    \Gamma_{\alpha,n}^\text{(All)} (E)\simeq \Big(K_{\alpha,n}^\text{(VPE)}+K_{\alpha,n}^\text{(NSpl)}\Big)\,\frac{G_F^2}{192\pi^3}\,E^{5+3n}\,\Dn^3\,,
    \label{eq:decaywidth_all}
\end{equation}
while, for energies $E\leq \Eth$, only the splitting contributes,
\begin{equation}
    \Gamma_{\alpha,n}^\text{(NSpl)} (E)=K_{\alpha,n}^\text{(NSpl)}\,\frac{G_F^2}{192\pi^3}\,E^{5+3n}\,\Dn^3\,,
    \label{eq:decaywidth_nspl}
\end{equation}

Note that for energies arbitrarily close to the VPE threshold eq.~\eqref{eq:decaywidth_all} is not exact, since it was obtained neglecting the electron mass. An estimate of the mass-dependent correction is
\begin{equation}
    \frac{m_e^2/E^2}{E^n\Dn}\sim \left(\frac{\Eth}{E}\right)^{n+2}\,.
    \label{eq:mass-correction}
\end{equation}
Naively, one could conclude from  eq.~\eqref{eq:mass-correction} that the correct expression of $\Gamma_{\alpha,n}$ reaches its asymptotic form already close to $E_n^{\text{th}}$. However, the estimate $(\Eth/E)^{n+2}$ only bounds the error from neglecting $m_e$ in the dispersion relation. In practice, the dominant suppression near threshold arises from the limited three-body phase space and the slow growth of its available volume, so that the asymptotic scaling is only recovered for energies 5 to 30 times $\Eth$, depending on the value of $n$. A detailed calculation of these threshold corrections lies beyond the scope of this work and will be presented elsewhere. In what follows, we will take eqs.~\eqref{eq:decaywidth_all} and \eqref{eq:decaywidth_nspl} as the expressions of the decay width, even if we know that we are overestimating it in an interval of energies above $\Eth$.

Finally, it is also worth to note that, even if the dispersion relations eq.~\eqref{MDRn0}, \eqref{MDRn12} are flavor independent, the dynamics of the decays depends on the flavor of the initial neutrino. The reason is that an electron neutrino can emit an $e^-e^+$ pair through both charged- and neutral-current interactions, while a muon or tau neutrino can only do so via the neutral current. This difference enhances the decay width for $\nu_e$ by a factor $\sim13$ (exactly $13$ when one takes $\sin^2\theta_W=1/4$) with respect to $\nu_\mu$ or $\nu_\tau$, as reflected in the coefficients of tables~\ref{table:Ks_VPE} and \ref{table:Ks_NSpl}. We can also note that a proper calculation of the decay width~\cite{Carmona:2012tp,Bezrukov:2011qn,Carmona:2022dtp} changes the coefficient $1/14$ indicated in ref.~\cite{Cohen:2011hx} to $17/420$ for muon and tau neutrinos, and to $221/420$, for electron neutrinos. 

\section{Constraints on superluminal neutrinos from the survival-probability method}
\label{sec:simplified}

We want to constrain superluminality from the observation of UHE neutrinos, which might be detected in the future with energies as high as $10^{19}\,\text{eV}$. In order to estimate the relevance of neutrino oscillations, let us note that the oscillation length increases with the energy, and for an energy around $10^{19}\,\text{eV}$, $L_\text{osc}\sim10^{17}\,\text{m}\sim10^{-3}\,\text{kpc}$. Assuming that the sources of UHE neutrinos are located beyond such a distance, a neutrino that reaches the Earth without decaying will, on average, oscillate many times during its propagation, spending about one third of the time in each flavour. It is then consistent to use a flavour-averaged total decay width to account for oscillation effects in the computation of the survival probability of traversing a distance $L\gg L_\text{osc}$. Let us define then $\Gamma_{n}^\text{(All)}$ and $\Gamma_{n}^\text{(NSpl)}$ as the result of using the flavour-averaged constants from tables~\ref{table:Ks_VPE} and \ref{table:Ks_NSpl} in eqs.~\eqref{eq:decaywidth_all} and \eqref{eq:decaywidth_nspl}, respectively.

The survival probability for propagation over a local (non-cosmological) distance $L$ is given by 
\begin{equation}
    \mathcal{P}^\text{SV}_n(E_d,L) = \exp\left(-L\,\Gamma_{n}(E_d)\right)\,.
    \label{eq:Psurvival}
\end{equation}
In this expression, $\Gamma_{n}$ refers to $\Gamma_{n}^\text{(All)}$ or $\Gamma_{n}^\text{(NSpl)}$ depending on whether the energy is above or below the threshold, respectively. We assume that trajectories with a survival probability $\mathcal{P}^\text{SV}_n < \epsilon$ are incompatible with a detection event during the detector's lifetime. Thus, a detection can be used to constrain the scale of new physics or the source distance. Following ref.~\cite{KM3NeT:2025mfl}, we take $\epsilon=\exp(-10)$ for the numerical calculations. Then, from the condition
\begin{equation}
    \mathcal{P}^\text{SV}_n(E_d,L) \geq \epsilon \,
\end{equation}
one gets that
\begin{equation}
    L \leq L_\text{max} \doteq \frac{\ln(1/\epsilon)}{\Gamma_n(E_d)} \,.
\end{equation}

Previous studies constraining superluminal neutrinos~\cite{KM3NeT:2025mfl,Yang:2025kfr,Satunin:2025uui} have neglected the flavor dependence of the decay widths. In addition, they often rely on the Cohen-Glashow approximation rather than on the full results of refs.~\cite{Carmona:2012tp,Bezrukov:2011qn,Carmona:2022dtp}, or restrict the analysis to the case $n=0$. In particular, ref.~\cite{KM3NeT:2025mfl} presented constraints on the LIV parameter $\delta$ for the $n=0$ case, employing the Cohen-Glashow coefficient $1/14 \approx 0.071$ instead of the correct value $17/84 \approx 0.202$ given in table~\ref{table:Ks_VPE}. In addition, the analysis does not incorporate the VPE threshold in their figure~1, effectively assuming $E > E^{\text{th}}_n$, while, as they point out in the main text, the validity of this assumption is limited. Constraints are then derived on $\Delta_0=\delta$ as a function of the source distance $L$, using the survival probability $ \mathcal{P}^\text{SV}_n$ introduced in eq.~\eqref{eq:Psurvival}, which strictly applies only to local sources. Nevertheless, this approximation is extended to cosmological distances in their figure~1, up to the scale of the observable Universe. In what follows, we address the shortcomings discussed above and develop a framework that overcomes these limitations. Our aim is to provide an analysis scheme that can be applied not only to the KM3-230213A event, but also to future detections of ultra-high-energy neutrinos.

For neutrinos emitted by extragalactic sources, the effect of cosmic expansion can no longer be neglected. In this case, the survival probability must account for the redshift evolution of the neutrino energy. Denoting by $E_d$ the detected energy and by $z$ the source redshift, eq.~\eqref{eq:Psurvival} is replaced by
\begin{equation}
    \mathcal{P}^\text{SV}_n(E_d,z) = \exp\left(-\int_0^{z} dz' \;\frac{\Gamma_{n}\big((1+z')E_d\big)}{(1+z')H_0\,h(z')} \right),
    \label{eq:Psurvivalexp}
\end{equation}
where $H_0$ is the value of the Hubble parameter at $z=0$ and $h(z)=\sqrt{\Omega_m (1+z)^3+\Omega_\Lambda}$. We consider an expansion described by the $\Lambda$CDM model, with $H_0=70\,\text{km/s/Mpc}$, $\Omega_m=0.3$ and $\Omega_\Lambda=0.7$. Assuming again that probabilities smaller than $\epsilon$ are incompatible with a detection event, one obtains a bound on the source redshift, $z \leq z_\text{max}$, where $z_\text{max}$ is defined as the solution of
\begin{equation}
     \int_0^{z_\text{max}} dz'\,\frac{(1+z')^{4+3n}}{h(z')} 
    \leq \frac{\ln(1/\epsilon) H_0}{\Gamma_n^\text{(All)}(E_d)}\,,
    \label{eq:boundaboveth}
\end{equation}
for energies above $\Eth$, and
\begin{equation}
     \int_{\Eth/E_d-1}^{z_\text{max}} dz'\,\frac{(1+z')^{4+3n}}{h(z')} 
    \leq \frac{\ln(1/\epsilon) H_0}{\Gamma_n^\text{(All)}(E_d)}-\frac{1}{1+K_{\alpha,n}^\text{(VPE)}/K_{\alpha,n}^\text{(NSpl)}} \int_0^{\Eth/E_d-1} dz'\,\frac{(1+z')^{4+3n}}{h(z')}\,,
    \label{eq:boundbelowth}
\end{equation}
for energies below the VPE threshold. Moreover, for a given source redshift $z$, these equations provide an upper bound on the LIV parameter $\Delta_n$, simply by replacing $z_\text{max}$ with the source redshift in eqs.~\eqref{eq:boundaboveth} and \eqref{eq:boundbelowth}.

\begin{figure}[p]
    \centering
    \includegraphics[height=8cm]{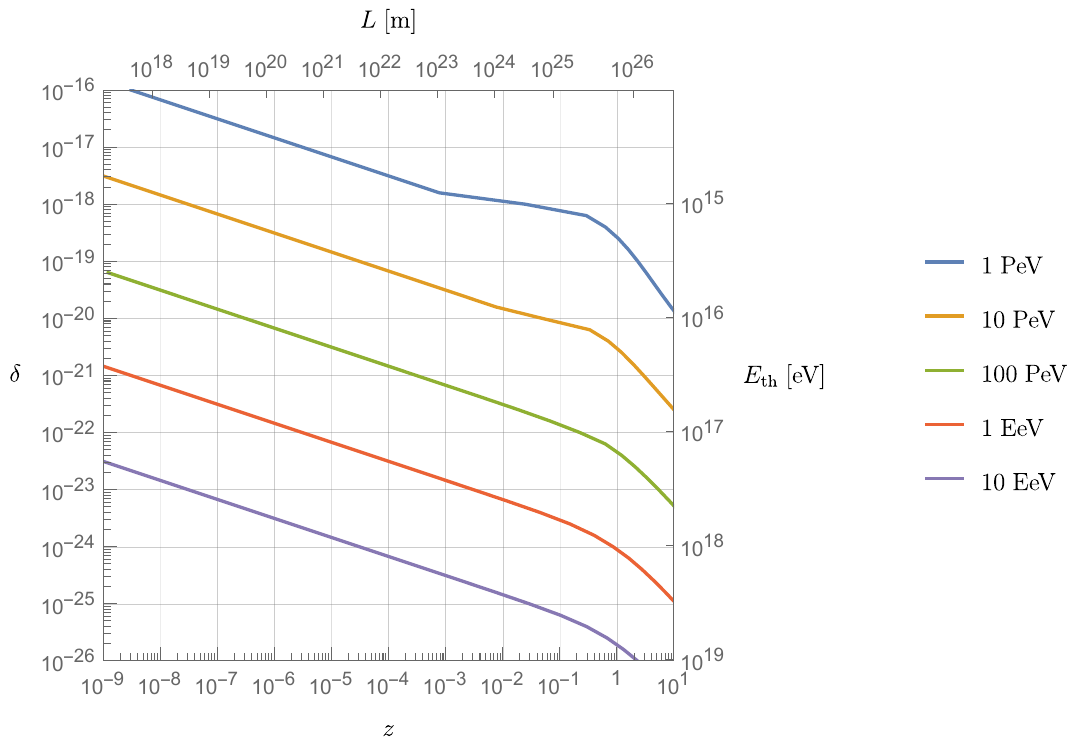}
    \caption{Constraints on the LIV parameter $\delta$ as a function of the source redshift $z$ (bottom axis) and the corresponding source distance $L$ in meters (top axis). 
    The left vertical axis shows the upper bound on $\delta$ obtained from eqs.~\eqref{eq:boundaboveth} and~\eqref{eq:boundbelowth} for different detected energies, each represented by a different color as indicated in the legend. 
    The right vertical axis displays the corresponding VPE threshold energy $E^\text{th}_{0}$ associated with the values of $\delta$ shown on the left axis.}
    \label{fig:n0curvas}
\end{figure}

\begin{figure}[p]
    \centering
    \includegraphics[height=8cm]{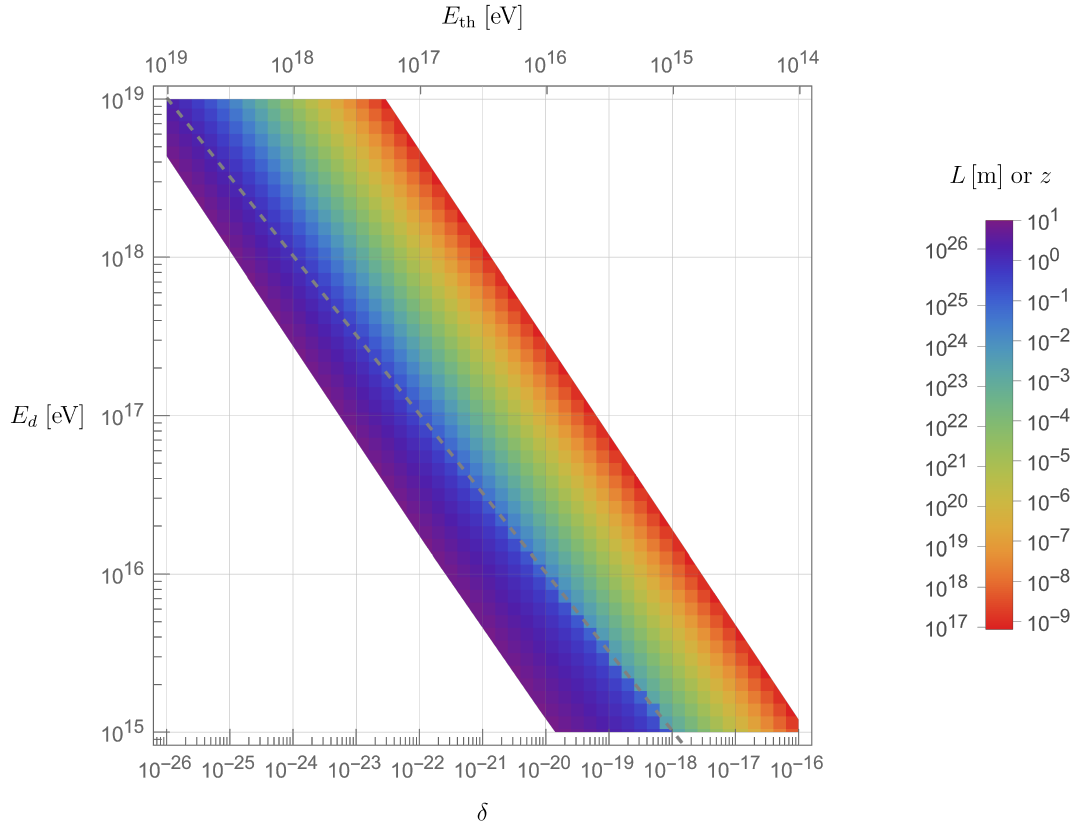}
    \caption{Heat-map representation of the constraints in the $(E_d,\delta,z)$ space for the $n=0$ case. The bottom axis shows the LIV parameter, while the top axis gives the corresponding VPE threshold energy, which is also shown as a dashed line in the plot. The left vertical axis corresponds to the detected energy $E_d$. Colors encode the constraint derived from the survival-probability condition in the range from $L=10^{17}\,\text{m}$ up to $z=10$. 
    A detected event corresponds to a horizontal line; its intersection with the colored boundary yields the upper bound on $\delta$ at each $z$.}
    \label{fig:n0heatmap}
\end{figure}

In the $n=0$ case, the curve in figure~1 of ref.~\cite{KM3NeT:2025mfl} is replaced by the set of upper-limit curves we show in our figure~\ref{fig:n0curvas}. These display the upper bound on $\delta$ as a function of the source redshift $z$ (bottom axis) and the corresponding comoving distance $L$ (top axis). The right vertical axis indicates the VPE threshold energy $E^\text{th}_{0}$ corresponding to the value of $\delta$ on the left vertical axis, and each colored curve corresponds to a different detected energy $E_d$. Values of $\delta$ above a given curve are excluded for a source at the corresponding $z$ and $E_d$. One can also see that the curves are no longer straight lines: the inclusion of cosmic expansion changes their slope at high redshift, and the resulting bound on $\delta$ becomes very strong. In fact, independently of the detected energy $E_d$, there always exists a sufficiently distant source ($z$ large enough) for which the survival probability amounts to $\epsilon$. The cosmological effect becomes important already at redshifts $z\sim 1$, which implies that neglecting it, as in ref.~\cite{KM3NeT:2025mfl}, leads to a physically relevant inconsistency in the resulting limits.

A more powerful way of visualizing the constraints is provided by the heat-map shown in figure~\ref{fig:n0heatmap}, where the three variables $(E_d,\delta,z)$ are combined into a single representation. Its key advantage is that a detected event corresponds simply to a horizontal line at the measured energy, so that the corresponding upper bound on $\delta$ can be directly read as a function of the source redshift $z$. This representation is therefore valid not only for the present analysis but also for future UHE neutrino detections. The plot spans from a source distance $L=10^{17}\,\text{m}$ (local Universe) up to a redshift $z=10$, and the dashed line indicates the VPE threshold energy associated with each value of $\delta$.

The same type of analysis can be extended to the quadratic case ($n=2$). 
The linear scenario ($n=1$) does not produce any bound in a LIV Lagrangian framework, since in that case neutrinos and antineutrinos propagate with opposite sign deviations from the speed of light~\cite{Carmona:2022dtp}. The constraints obtained by ref.~\cite{Satunin:2025uui} in the $n=1$ case would then only be valid beyond such a framework. 

\begin{figure}[p]
    \centering
    \includegraphics[height=8cm]{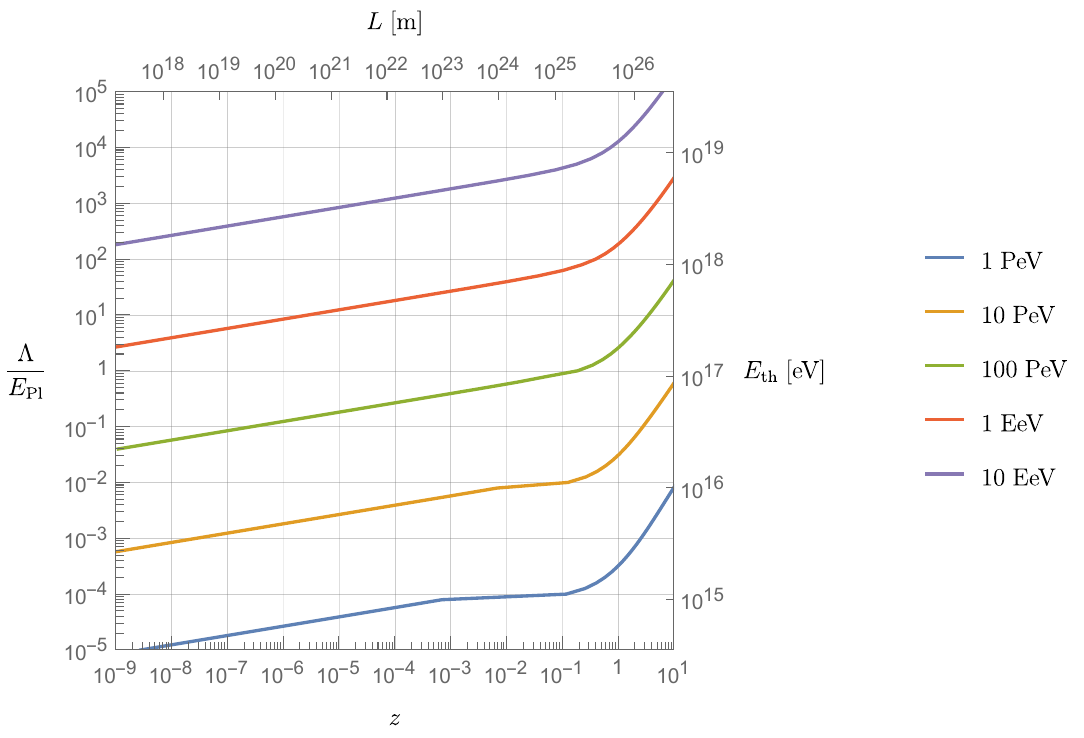}
    \caption{Constraints on the LIV parameter $\Lambda$ in units of the Planck energy $E_\text{Pl}$ as a function of the source redshift $z$ (bottom axis) and the corresponding source distance $L$ (top axis), for fixed detected energy $E_d$. The left vertical axis shows the bound on $\Lambda/E_\text{Pl}$ obtained from the survival-probability analysis, while the right vertical axis indicates the associated VPE threshold energy. As in the $n=0$ case, the inclusion of cosmic expansion modifies the slope of the curves at large $z$, and the bound on $\Lambda$ becomes significantly stronger.}
    \label{fig:n2curvas}
\end{figure}

For $n=2$, the relevant parameter to be constrained is not $\delta$, but the LIV scale $\Lambda$, which we express in units of the Planck energy $E_\text{Pl}$. Figure~\ref{fig:n2curvas} shows the lower bound on $\Lambda/E_\text{Pl}$ as a function of the source redshift $z$ (bottom axis) and the corresponding source distance $L$ (top axis), for a fixed detected energy $E_d$. As in the $n=0$ case, the curves deviate from straight lines at high redshift due to the effect of cosmic expansion, which substantially tightens the constraint. 

\begin{figure}[p]
    \centering
    \includegraphics[height=8cm]{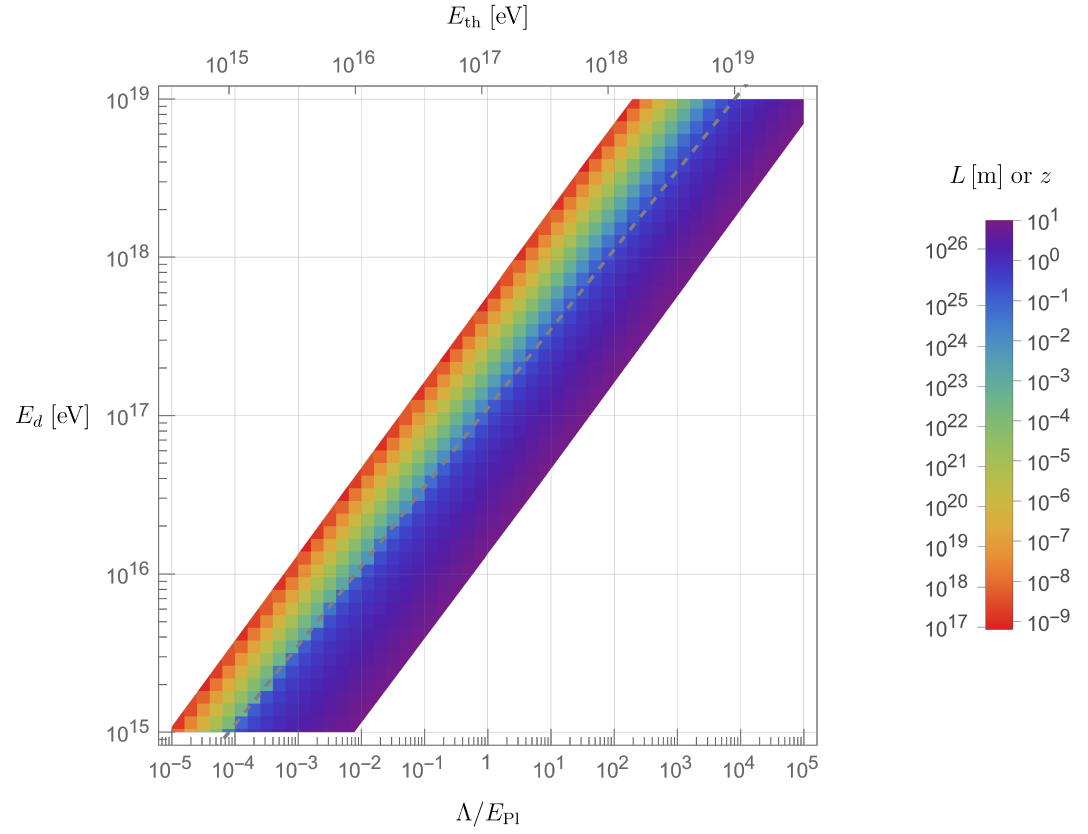}
    \caption{Heat-map representation of the constraints in the $(E_d,\Lambda/E_\text{Pl},z)$ space for the $n=2$ case. The bottom axis shows the LIV parameter, while the top axis gives the corresponding VPE threshold energy, which is also shown as a dashed line in the plot. The left vertical axis corresponds to the detected energy $E_d$. Colors encode the constraint derived from the survival-probability condition in the range from $L=10^{17}\,\text{m}$ up to $z=10$. 
    A detected event corresponds to a horizontal line; its intersection with the colored boundary yields the lower bound on $\Lambda$ at each $z$.}
        \label{fig:n2heatmap}
\end{figure}

A more event-independent view is provided in figure~\ref{fig:n2heatmap}, which displays a heat-map representation in the $(E_d,\Lambda/E_\text{Pl},z)$ space. 
A detected event of energy $E_d$ corresponds to a horizontal line, and the intersection with the shaded region directly provides the lower bound on $\Lambda/E_\text{Pl}$ as a function of the source redshift. 
The plot spans the range $L=10^{17}\,\text{m}$ to $z=10$, and the dashed curve marks the corresponding threshold energy. 

\section{Validity of the survival-probability approximation}
\label{sec:cascades}

The analysis carried out so far has been implicitly based on the assumption that the detected neutrino was directly emitted by the source. In other words, the survival-probability formalism accounts only for the case in which the neutrino has propagated from the source without undergoing any decay. However, in the presence of superluminal propagation, neutrinos may decay along their trajectory, giving rise to a cascade of secondary neutrinos that can also be detected.

In order to clarify this point, let us first recall that the flux of detected neutrinos with energy $E_d$ after propagation from a source at a (local) distance $L$ can be written as
\begin{equation}
    \phi_d^{(0)}(E_d, L) = \phi_e(E_d)\, \mathcal{P}_n^\text{SV}(E_d,L)= \phi_e(E_d)\, \exp\left(-L\,\Gamma_n(E_d)\right) \,,
\end{equation}
where $\phi_e(E_d)$ denotes the spectral emission distribution at the source and the exponential factor represents the survival probability assuming no decay along the path. We drop the subindex $n$ from now on to easy the notation.

One must also take into account the possibility that the detected neutrino originates from a \emph{cascade}. For instance, the flux that results from a \emph{single decay} occurring at a distance $xL$ from the source is given by
\begin{equation}
\label{eq:P1}
    \phi_d^{(1)}(E_d, L) = \int_0^1 dy \, f(y)\, \phi_e\left(E_d/y\right) 
    \int_0^1 e^{-(1-x)L\,\Gamma(E_d/y)} \, L\,dx\, \Gamma(E_d/y)\, e^{-xL\,\Gamma(E_d)}\,,
\end{equation}
where $f(y)$ is the distribution of the energy fraction $y$ carried by the secondary neutrino produced in the decay. The first exponential factor accounts for the survival probability of the parent neutrino up to the decay point, the middle factor gives the decay probability within a distance interval $L\,dx$, and the last exponential factor represents the survival of the daughter neutrino on its way to the detector.

The general case corresponds to the sum over an arbitrary number of decays along the propagation, so that the total detected flux from a source at distance $L$ is obtained as
\begin{equation}
    \phi_d(E_d,L) = \sum_{k=0}^\infty \phi_d^{(k)}(E_d,L) \,.
\end{equation}

The quotient between the detected and emitted flux at the same energy, provides an appropriate generalization for the probability of survival at local distances,
\begin{equation}
    \frac{\phi_d(E_d,L)}{\phi_e(E_d)} = \sum_{k=0}^\infty \frac{\phi_d^{(k)}(E_d,L)}{\phi_e(E_d)} \doteq \sum_{k=0}^\infty \mathcal{R}^{(k)} \,.
\end{equation}
Note that for $k=0$, $\mathcal{R}^{(0)}=P_n^\text{SV}(E_d,L)$, but for $k>0$ the quantities $\mathcal{R}^{(k)}$ are not normalized probabilities but contributions to the spectral flux at energy $E_d$.

The simplified method introduced in section~\ref{sec:simplified}, which retains only $\mathcal{R}^{(0)}$, can in principle underestimate the flux of cascade neutrinos. To assess the importance of this effect, let us consider the contribution of a single decay, ${\cal R}^{(1)}$, in the simplest case, $n=0$. Since for $n=0$ only vacuum $e^-e^+$ emission is allowed, we will only consider energies $E_d>E_0^\textrm{th}$. If $E_d<E_{0}^\textrm{th}$ the decay width vanishes and no constraint on LIV can be derived.

The integral over $x$ in eq.~\eqref{eq:P1} can be easy solved, 
\begin{equation}
    \int_0^1 dx \,e^{x\,L\, [\Gamma(E_d/y) - \Gamma(E_d)]} \,=\, \frac{ e^{L\, [\Gamma(E_d/y) - \Gamma(E_d)]}-1}{L\, [\Gamma(E_d/y) - \Gamma(E_d)]}\,.
\end{equation}
Replacing the result in eq.~\eqref{eq:P1} one obtains
\begin{equation}
\phi_d^{(1)}(E_d, L) \,=\, \int_0^1 dy \, f(y)\, \phi_e(E_d/y)\, \frac{e^{-L\,\Gamma(E_d)} - e^{-L\,\Gamma(E_d/y)}}{1-\Gamma(E_d)/\Gamma(E_d/y)}\,.
\end{equation}
Using the energy dependence of the decay width, eq.~\eqref{eq:decaywidth_all}, and assuming an emission spectrum $\phi_e(E) \propto E^{-\gamma}$, one readily obtains
\begin{equation}
    \Gamma(E_d/y) \,=\, (1/y^5)\, \Gamma(E_d), \hskip 1cm \phi_e(E_d/y) \,=\, y^\gamma\, \phi_e(E_d)\,.
\end{equation}
Thus,
\begin{equation}
    \mathcal{R}^{(1)} = \frac{\phi_d^{(1)}(E_d,L)}{\phi_e(E_d)} \,=\,e^{-L\Gamma(E_d)} \int_{0}^1 dy\, f(y)\, y^\gamma\, \left[\frac{1-e^{-L\Gamma(E_d) (1/y^5 - 1)}}{1-y^5}\right].
\end{equation}

It is clear that $\mathcal{R}^{(1)}$ is upper bounded by 
\begin{equation}
    \mathcal{R}^{(1)} < e^{-L\Gamma(E_d)} \int_{0}^1 dy\, f(y)\, \frac{y^\gamma}{1-y^5} = e^{-L\Gamma(E_d)} \left<\frac{y^\gamma}{1-y^5}\right> \,.
\end{equation}
Higher-order contributions would involve even larger parent energies, where the source emission is strongly suppressed. In fact, from a general study of the contributions in the cascade, it is possible to show that
\begin{equation}
    \mathcal{R}^{(k)} < e^{-L\Gamma(E_d)} \left<\frac{y^\gamma}{1-y^5}\right>^k \,,
\end{equation}
for every value of $k$.\footnote{This analysis lies beyond the scope of the present work and will be presented elsewhere.}
Then, a bound on the quotient between the detected and emitted flux can be build as a geometric sum,
\begin{equation}
    \frac{\phi_d(E_d,L)}{\phi_e(E_d)} < e^{-L\Gamma(E_d)} \,\sum_{k=0}^\infty \left<\frac{y^\gamma}{1-y^5}\right>^k = e^{-L\Gamma(E_d)}\,\left(1-\left<\frac{y^\gamma}{1-y^5}\right>\right)^{-1}\,.
\end{equation}

Under the assumption that the detected particle is a primary, we rejected the possibility of detecting a neutrino with a probability of survival smaller than $\epsilon$. If we now consider the possibility that the particle can be produced inside the cascade, we should now reject the possibility of detecting a neutrino of energy $E_d$ if $\phi_d(E_d,L)/\phi_e(E_d)<\epsilon$. Then, a sufficient condition for rejection is 
\begin{equation}
     \frac{\phi_d(E_d,L)}{\phi_e(E_d)} < e^{-L\Gamma(E_d)}\,\left(1-\left<\frac{y^\gamma}{1-y^5}\right>\right)^{-1} < \epsilon \,.
\end{equation}
We see that this new condition is equivalent to change the value of $\epsilon$ by a factor $\left[1-\left<y^\gamma/(1-y^5)\right>\right]$. One can explicitly see in this factor the competition between spectral suppression (the factor $y^\gamma$) and the dynamical enhancement due to the faster decay of the parent (the factor $1-y^5$ in the denominator).

For $n=0$ and $\gamma\sim 2$, we find $\left<y^\gamma/(1-y^5)\right>\approx 0.063$, which shows that the correction induced by cascade regeneration is minor (at most a few percent). Given this small correction, combined with the arbitrariness on the choice of $\epsilon$, we conclude that cascade effects can be safely neglected when setting LIV bounds using the $\phi_d(E_d,L)/\phi_e(E_d) < \epsilon$ rejection criterion, and the simpler survival-probability approximation is therefore justified. However, this conclusion is strictly limited to the ``cut-off'' regime, i.e., when the flux ratio $\phi_d(E_d,L)/\phi_e(E_d)$ is very small. In situations where neutrino propagation is more probable ($L\Gamma(E_d)\lesssim 1$) --- such as at lower energies or shorter baselines --- cascade regeneration may play a significant role and must be explicitly included in the analysis. This effect is expected to be even more relevant in the $n=2$ case, which produces additional secondary neutrinos. Even so, this does not change our qualitative conclusion regarding the robustness of the methodology used in the section~\ref{sec:simplified} and in ref.~\cite{KM3NeT:2025mfl}.

\section{Time delays, KM3-230213A and future UHE neutrino events}
\label{sec:KM3future}

The unified framework developed in section~\ref{sec:thdecay} allows one to derive updated constraints on LIV from the detection of ultra-high-energy neutrinos. In this section we apply these results to the specific case of the KM3-230213A event recently reported by the KM3NeT collaboration, and we discuss their implications for present and future observations.

\subsection{Constraints from the KM3-230213A event}

The KM3-230213A event corresponds to a neutrino of energy $220^{+570}_{-100}$\,PeV. Using the formalism of section~\ref{sec:simplified} and the results in figures~\ref{fig:n0curvas}, \ref{fig:n0heatmap}, \ref{fig:n2curvas} and \ref{fig:n2heatmap}, we translate this detection into constraints on superluminal LIV for both the energy-independent ($n=0$) and quadratic ($n=2$) scenarios. To derive the most conservative constraints, we adopt the lower bound of the event's energy interval, approximated here as $E_d \approx 100$\,PeV.

\begin{itemize}

\item Case $n=0$:
For sources in the local Universe the bounds on $\delta$ are very close to those reported by KM3NeT: the slope of the constraint as a function of distance/redshift is practically the same, with only a modest ($\sim$20\%) tightening attributable to the use of the full decay widths instead of the Cohen-Glashow approximation and the proper flavour averaging. The most relevant difference appears for cosmological sources, where the expansion is treated consistently here: the curves bend down at large redshift (cf.\ figure~\ref{fig:n0curvas}), leading to \emph{stronger} limits than in ref.~\cite{KM3NeT:2025mfl}. In particular, for $z \in [1,3]$ we obtain
\[
\delta \;\lesssim\; 10^{-23}\text{--}10^{-22},
\]
whereas the KM3NeT estimate stays closer to $\delta \sim 10^{-22}$ because the high-$z$ propagation is not handled consistently in their survival-probability treatment. This difference is precisely the effect of the cosmological redshifting of the energy inside the decay width.

\item Case $n=2$:
For the quadratic case, the constrained parameter is the LIV scale $\Lambda$ defined by $E^2 = p^2[1+(p/\Lambda)^2]$. From figures.~\ref{fig:n2curvas} and \ref{fig:n2heatmap} we find
\[
\Lambda \;\gtrsim\; (3\times 10^{-2}\text{--}5\times 10^{1})\,E_{\rm Pl},
\]
with the precise value depending on the assumed source redshift (higher $z$ implies stronger bounds due to cosmological expansion). These constraints provide the current limits on superluminal neutrino propagation consistent with the observed stability of the KM3-230213A event, and include both VPE and NSpl, their thresholds, and the proper cosmological propagation.

\end{itemize}

\subsection{Compatibility with time-of-flight effects}

The implications of Lorentz violation in decay kinematics are closely tied to its possible manifestations in time-of-flight effects. A superluminal dispersion relation implies both (i) a finite decay width at high energy, which threatens the stability of UHE neutrinos over cosmological distances, and (ii) an advance in the arrival time with respect to photons or gravitons from the same transient source. These two aspects have so far been probed separately: on the one hand, searches for neutrino-photon temporal correlations in association with gamma-ray bursts and other transients constrain possible time-of-flight advances; while on the other hand stability arguments applied to the recent KM3-230213A event constrain the allowed decay width at $\sim 100$\,PeV. As emphasized in ref.~\cite{Carmona:2023mzs}, a combined analysis of both effects for the same event would provide a powerful internal consistency test of LIV.
Following this reference, the arrival-time advance relative to a photon can be written, as
\begin{align}
 \Delta t_n (E_d,z) = \frac{n+1}{2H_0}\,E_d^n\,\Delta_n \int_0^{z} dz' \frac{(1+z')^n}{h(z')} \,.
\end{align}
$H_0^{-1}\approx4.4\times10^{17}\,$s is the Hubble time, and the redshift integral is of order unity for $z\!\sim\!1$, reaching values $\sim8$ for $n=2$  and $z\!\sim\!3$~\cite{Carmona:2023mzs}.

The bounds over the scale of new physics obtained in the previous section from the stability analysis, can be translated to constraints for the time delay. Taking $E_d=100\,$PeV as before, we have:
\begin{itemize}

\item Case $n=0$:  
For $\delta\!\sim\!10^{-23}$--$10^{-19}$, one finds
\begin{equation}
\Delta t_{0} \lesssim (10^{-6}\text{--}10^{-2})~\text{s},
\end{equation}
ranging from microseconds to tens of milliseconds. In this scenario, the time advance $\Delta t_{0}$ is energy independent, whereas the decay width grows steeply with $E_d$. Consequently, stability arguments provide the most stringent bounds at high energies, while time-of-flight sensitivity does not improve. The resulting upper limits on $\delta$ thus imply that any superluminal advance for UHE neutrinos must be sub-second. Although high-energy time delay measurements are not enhanced, timing observations at lower energies (e.g., sub-PeV) serve as a crucial consistency test: the absence of a measurable delay is \emph{required} for the stability-based limits derived here to be physically self-consistent. A significant deviation at low energies would point to a LIV scenario beyond the EFT framework used to compute the decay widths. Finally, should future multimessenger observations detect a significant positive time delay (neutrinos arriving later than photons), this would favor a subluminal interpretation and effectively exclude the superluminal one.

\item Case $n=2$:  
For $\Lambda\!\sim\!(3\times10^{-2}\text{--}5\times10^{1})E_{\rm Pl}$, the associated time advance is
\begin{equation}
   \Delta t_{2} \lesssim (2\times 10^{-8}\text{--}4\times 10^{-1})\ \text{s}, 
\end{equation}
depending on $\Lambda$ and $z$. In this scenario, both the decay width and the time advance scale as $E^2/\Lambda^2$.  
As a result, the stability of the $\approx 100\,\mathrm{PeV}$ KM3--230213A neutrino already enforces that any superluminal time advance remains very small. Although increasing the neutrino energy would enhance $\Delta t_{2}$ for a fixed $\Lambda$, the stability requirement simultaneously pushes the allowed values of $\Lambda$ upward, so that the maximal compatible time advance remains at the sub-second level. This implies that, as in the $n=0$ case, timing measurements at lower energies already provide a crucial cross-check: their consistency with null time delays would confirm that the same LIV parameters governing decay kinematics also control propagation.

\end{itemize}

A mismatch between stability and timing constraints—namely, a significant time delay incompatible with the decay-based bounds—would signal a departure from the EFT description of LIV.  
In such a case, the linear scenario ($n=1$), where neutrinos and antineutrinos receive LIV corrections of opposite sign, would become particularly relevant: if the underlying dynamics lies beyond the EFT regime, the symmetry between superluminal and subluminal behavior could be broken, giving rise to effects distinguishable from those of the standard framework.

In summary, the KM3-230213A event already constrains superluminal propagation to a level where any associated time-of-flight effect is expected to lie well below present detection thresholds.  
Future UHE neutrino observations, especially in combination with precise timing studies of lower-energy events, will provide a stringent internal consistency test of Lorentz invariance, probing whether the underlying mechanism conforms to an effective field theory description or requires a more fundamental extension.

\section{Conclusions and future prospects}
\label{sec:conclusions}

In this work we have revisited the constraints on Lorentz Invariance Violation (LIV) that can be derived from the detection of ultra-high-energy (UHE) neutrinos, using as reference the KM3-230213A event recently observed by the KM3NeT collaboration. We have presented a unified and self-consistent framework for superluminal neutrino decays, valid for both the energy-independent ($n=0$) and quadratic ($n=2$) dispersion-relation scenarios. The corresponding decay widths and threshold energies have been consistently implemented, together with the proper cosmological propagation. We have shown that the simplified method based on the survival probability provides a robust approximation for setting LIV bounds, since cascade regeneration effects are suppressed by the steep energy dependence of the decay width and the small survival probabilities used in the limit extraction.

The formalism developed here can be directly applied to future UHE neutrino detections, and it leads to constraints on the LIV parameters that are driven only by the survival probability of an observed neutrino of energy $E_d$ over a known (or assumed) source distance. In this sense, the bounds on $\delta$ (for $n=0$) or on $\Lambda$ (for $n=2$) obtained from eqs.~\eqref{eq:boundaboveth} and~\eqref{eq:boundbelowth} do not depend on the detailed shape of the source spectrum: they follow from the requirement that at least one neutrino with energy $E_d$ can reach Earth without decaying with probability larger than $\epsilon$. 

By contrast, the source spectrum does play an important role when going beyond the existence of a single detected event and addressing the predicted flux. In particular, it controls the relative weight of regenerated neutrinos produced in superluminal cascades, and it determines the expected cosmogenic contribution at the highest energies. Thus, while the LIV bounds themselves are largely spectrum-independent, any attempt to model the population of UHE neutrinos and its origin (e.g. cosmogenic vs.\ source-accelerated) will require assumptions about the injection spectrum, spectral index, and high-energy cutoff.

A second important aspect is the possible distinction between neutrinos and antineutrinos at ultra-high energies. The linear case ($n=1$) differs qualitatively from the $n=0$ and $n=2$ scenarios, since for $n=1$ the sign of the LIV parameter alternates between particles and antiparticles: superluminal neutrinos imply subluminal antineutrinos and vice versa. Determining whether the detected UHE flux is dominated by neutrinos, antineutrinos, or a mixture of both will thus be crucial to extend these analyses to the linear LIV case.

This will be relevant for scenarios of LIV violation which go beyond the standard EFT framework. Since the same LIV parameters that control the stability of UHE neutrinos also determine potential deviations in their time of flight with respect to photons or gravitons from the same transient source, any incompatibility between constraints from time delays and neutrino stability may point to such scenarios.

The detection of UHE neutrinos opens a powerful avenue to test Lorentz invariance at unprecedented energies. The next years will likely bring multiple detections of UHE neutrino events by both IceCube and KM3NeT, offering a unique opportunity to cross-validate of energy scales, flavor composition, and sky coverage. A combined analysis of their observations could significantly improve the sensitivity to LIV, as well as to the spectral and directional properties of the UHE neutrino population.
The present study establishes a consistent foundation for these analyses, demonstrating that the survival-probability approximation is adequate for setting current bounds, while also outlining the conditions under which cascade effects, flavor composition, or spectral uncertainties may become significant in future, more precise observations.

\section*{Acknowledgments}

This work is supported by the Spanish grants PGC2022-126078NB-C21 and PID2024-160228NB-I00, funded by MCIN/\-AEI/\-10.13039/\-501100011033 and `ERDF A way of making Europe’, grant E21\_23R funded by the Aragon Government and the European Union, and the NextGenerationEU Recovery and Resilience Program on `Astrofísica y Física de Altas Energías’ CEFCA-CAPA-ITAINNOVA. The authors would like to acknowledge the contribution of the COST Actions CA18108 ``Quantum gravity phenomenology in the multi-messenger approach'' and CA23130 ``Bridging high and low energies in search of quantum gravity''.

\bibliographystyle{apsrev4-1}
\bibliography{main}

\end{document}